\documentclass[a4paper,fleqn]{cas-sc}

\usepackage[authoryear,longnamesfirst]{natbib}

\usepackage{amsmath}
\usepackage{nccmath}
\DeclareUnicodeCharacter{2264}{ }
\DeclareUnicodeCharacter{2295}{ }
\DeclareUnicodeCharacter{2212}{ }
\DeclareUnicodeCharacter{2217}{ }

\ExplSyntaxOn
\keys_set:nn { stm / mktitle } { nologo }
\ExplSyntaxOff

\begin{document}
\let\WriteBookmarks\relax
\def\floatpagepagefraction{1}
\def\textpagefraction{.001}
\shorttitle{M. I. Saeed et~al./New Astronomy (2020)}
\shortauthors{M. I. Saeed et~al.}

\title [mode = title]{Multi-color photometry and parameters estimation of Jupiter-sized exoplanets; TRES-3b, WASP-2b and HATP-30b}

\author[1]{M. I. Saeed}
\cormark[1]
\fnmark[1]
\ead{irfan.saeed22@gmail.com}


\address[1]{Institute of Space Technology, Islamabad, 44000, Pakistan}

\author[2]{S. N. Goderya}

\fnmark[2]
\ead{goderya@tarleton.edu}


\address[2]{Department of Chemistry, Geoscience and Physics, Tarleton State University, Stephenville, TX 76008 USA}

\author[3] {F. A. Chishtie}

\fnmark[3]
\ead{fachisht@uwo.ca}

\address[3]{Department of Physics and Astronomy, University of  Western Ontario, London, ON N6A 3K7, Canada}

\cortext[cor1]{Corresponding author}

\begin{abstract}
Precise and frequent photometric follow-up studies of transit light curves are indispensable when accurately characterizing extrasolar planets. We present new multi-wavelength photometry of three transiting "Hot Jupiters" — TrES-3b, WASP-2b and HAT-P-30b (WASP-51b). Data were acquired from an 0.8 meter telescope at Tarleton State University. When combined with literature data, allowed us to redetermine system parameters in a corresponding way. We developed GCX reduction pipeline and TAP - modeling and light curve fitting package to analyze the extracted light curves. We then used weighted mean results to estimate the parameters from BVRI filters for three exoplanetary systems and compared them to previous results. We concluded our determined parameters are agreed with previous studies. From our study, TreS-3b with a mass of $M_p$ = 1.773$M_{jup}$ and $R_p$ = 1.305$R_{jup}$ appears slightly less massive, while HAT-P-30b, which has a mass of $M_p$ = 0.7006$M_{jup}$ and $R_p$ = 1.5109$R_{jup}$ appears to be a bloated "Hot Jupiter". Additionally, we compared the results of our broadband photmetric analysis with the previous studies to search for transit depth wavelength dependence. We found a flat spectrum across optical wavelengths (except WASP-2b in R-band) for TrES-3b and WASP-2b, indicating the presence of clouds in their atmospheres. For WASP-2b, $R_p$/$R_*$ value was 0.14215 which was 0.14$\sigma$ higher than in the previous work. HAT-P-30b had a significantly larger radius in B filter with $R_p$ = 0.161246$R_{jup}$, resulting from $R_p$/$R_*$ = 0.1334, which was secondarily confirmed by the atmospheric scale height value, H = 1450km, indicating that HAT-P-30b is an inflated "Hot Jupiter".

\end{abstract}


\begin{highlights}
\item We present multi-wavelength light curves of TrES-3b, WASP-2b and HAT-P-30b
\item GCX (Exoplanet data reduction pipeline) was developed.
\item TAP, Transit Analysis Package, was also used for this project.
\end{highlights}


\maketitle

\section{Introduction}

\paragraph{} Discovery of the first exoplanet using radial velocity method (RV) by \citet{mayor1995jupiter} is one of the greatest scientific achievements of the 20th century. However, radial velocity data does not provide detailed parameters of exoplanets orbiting their host stars. As it can only determine the planet's eccentricity (e), orbital period (P), planet to host star separation distance or semi-major axis (a) and lower mass limit of the planet \cite{southworth2010homogeneous}. This problem was resolved for the first time when an exoplanet HD209458 was discovered by \citet{charbonneau1999detection} traversing in front of its host star and was dimming a very minute amount of light coming from its host star, the method known as transit method. Transit method is the only method that allows the direct measurement of planetary radius and thus for the determination of other planetary parameters such as density, gravity, chemical composition, semi-major axis and eccentricity \cite{charbonneau2007precise}. It also allows us to confront the observations with the existing theoretical models. The basic astrophysical parameters of planetary systems determined by this technique are more precise than obtained by other methods. (e.g. \cite{southworth2008homogeneous}; \cite{torres2008improved}; \cite{southworth2012homogeneous}; \cite{ciceri2015physical}). The absorption properties of different species in a planetary atmosphere vary with wavelength, causing an observable variation in the planet's radius. Multi-band photometry of a transiting exoplanet (TEP) system can be used to constrain the composition of an exoplanet's atmosphere (\cite{seager2000theoretical}; \cite{brown2001transmission}; \cite{charbonneau2002detection}). For transiting planets, there is a unique opportunity to determine the mass and radius, and hence the mean density, which is a key parameter for studies of planetary internal structure. Modeling can then constrain the chemical composition and the mass of the solid core \cite{nettelmann2010interior}. It can also further help in understanding the formation and evolution of stellar and planetary systems. Most of these TEPs have been found in large-scale transit surveys such as Kepler \cite{borucki2010kepler}, K2 \cite{howell2014k2}, WASP \cite{pollacco2006wasp}. Ground-based photometric surveys along with space missions play an important role in discovering hot Jupiters. Follow-up studies by ground-based telescopes are helpful in detailed parametric studies and are then archived in an online database exoplanet transit database by many observers \cite{poddany2010exoplanet}.

\paragraph{} TrES-3b is a hot Jupiter orbiting G-type dwarf star ($mass = 0.928 \pm 0.038M_{sun}$, $radius = 0.829 ± 0.02R_{sun}$ and metallicity $(Fe/H) = -0.19 ± 0.08)$ with an orbital period of $1.30619 ± 0.00001$ days, $mass = 1.92 ± 0.23M_{jup}$ and $radius = 1.295 ± 0.081R_{jup}$ \cite{o2007tres}. The planet with its almost grazing orbit was detected by two different transit surveys – the Trans-atlantic Exoplanet Survey (TrES) and the Hungarian Automated Telescope Network (HATNet). Several follow-up primary transit photometric studies have confirmed these planetary parameters (e.g. \cite{gibson2009transit}; \cite{sozzetti2009new}; \cite{colon2010characterizing}; \cite{ballard2011kepler}. \cite{sozzetti2009new}) precisely measured the astrophysical parameters of the TrES-3 system by using new photometric and spectroscopic data and found deviations in period of TrES-3b due to presence of orbiting bodies in the system. \citet{gibson2009transit} carried out nine follow-up photometric observations, combined with the previous data, to search for the additional planet in the system and predicted the presence of earth-mass planets in circular orbit, present at 2:1 mean resonance of TrES-3b. \citet{christiansen2010system} suggested this transit timing variation (TTV) due to the presence of star spots. However, \citet{lee2011physical} and \citet{turner2013near} nullified the periodical variation in TTV by suggesting the activities due to stellar magnetic fields. \citet{jiang2013possible} suggested the existence of periodic TTV with single frequency. \citet{vavnko2013photometric} rejected the possibility of the presence of periodic TTV by analyzing many observations along with data from other literature.

\paragraph{} WASP-2b was discovered by \cite{cameron2007wasp}\cite{wilson2007superwasp} in photometry from the SuperWASP-North telescope \cite{pollacco2006wasp} and and the Wide Angle Search for Planets (WASP) project \cite{enoch2011wasp}. WASP-2b lies in a category between hot Jupiters and very hot Jupiters, has mass = 0.85$M_{jup}$ and radius = 1.02$R_{jup}$ orbiting around K1 host star (V=12) with approx. 2.15days period. Planet is slightly larger and less massive than Jupiter while star WASP-2A is smaller and cooler than the Sun. WASP-2b orbits its host star at minimum separation below which mass-loss activity could dominate. Follow-up photometry of one transit was presented and analyzed by \cite{charbonneau2007precise}, and reanalyzed by \cite{southworth2008homogeneous}\cite{southworth2009homogeneous}.

\paragraph{} HAT-P-30b (also known as WASP-51b) was independently discovered by both the Hungarian-made Automatic Telescope Network (HAT-Net) \cite{johnson2011hat} and the Wide Angle Search for Planets (WASP) project \cite{enoch2011wasp}. According to \cite{johnson2011hat}, it has a $radius = 1.340 ± 0.065R_{jup}$ (Jupiter radius) and $mass = 0.711 ± 0.028M_{jup}$ (Jupiter mass) that gives a relatively low mean density of $/rho_p = 0.279 ± 0.038rho_{jup}$ (Jupiter density). Its orbital period is approx. 2.81 days. The host star BD+06 1909 (V =10.35 mag) is a one Gyr-old dwarf of spectral type F. All the information related to the magnitude and location of stars in sky is provided in table 2.\\

\begin{table}
\caption{Parameters derived from spectroscopic data of TrES-3, WASP-2 and HAT-O-30 stars used in our model fitting of light-curves to calculate the limb darkening coefficients and weighted-mean parameter values.}\label{tbl1}
    \centering
    \begin{tabular}{ p{3cm}||p{3cm}p{3cm}|p{3cm} }
\hline
Parameters& TrES-3& WASP-2& HAT-P-30\\
\hline

Mass M*($M_{sun}$) &0.928±0.035 &0.79±0.1 &1.242±0.041\\
Radius R*($R_{sun}$) &0.829±0.02 &0.78±0.06 &1.215±0.051\\
Density (cgs)& 2.304±0.066 &1.012±0.09& -\\
Surface gravity (logg*)(cgs)& 4.4±0.1& 4.5±0.3& 4.36±0.03\\
Eff. temp. ($T_{eff}$)& 5650±75& 5200±200& 6304±88\\
Fe/H& −0.19±0.08& 0.1±0.2& 0.13±0.08\\
RV ($K_*$)(m/s)& 369.8±7.1& 155±7& 88.1±3.3\\
$vsini$ (km/s)& 1.5±1.0& 0.99& 2.2±0.5\\
$M_v$(mag)& 5.39±0.11& 6.2±0.5& 3.98±0.14\\
Distance(pc)& 228±12& - & 193±\\
Age& 0.9+2.8−0.8& - & 1.0+0.8−0.5\\
Reference& \citet{sozzetti2009new}& \citet{cameron2007wasp}& \citet{johnson2011hat}\\
\hline
\end{tabular}
\end{table}

\begin{table}
\caption{Details of host stars of planetary systems studied in our work}\label{tbl1}
    \centering
    \begin{tabular}{ c||c|c|c }
\hline
Star& RA(J2000)& Dec(J2000)& Vmag\\
\hline
TrES-3&                 17 52 07.02& 37 32 46.2& 12.2\\
WASP-2&                 20 30 54.14& 06 25 46.7& 12.1\\
HAT-P-30/WASP-51&    08 15 47.98& 05 50 12.4& 10.4\\
\hline
\end{tabular}\\
\end{table}

\section{Observations and Data Reduction} 

\paragraph{1. Data from literature} The parameters were published in literature and these are determined from spectroscopic radial velocity data. These parameters were incorporated in our current analysis (see table 1). For TrES-3b study, we included High-Resolution Echelle Spectrometer (HIRES) on the Keck I telescope (HIRES;\cite{vogt1994hires}) RV data from \citet{sozzetti2009new} with spectroscopic parameters $T_{eff}$ = 5650.0K, [Fe/H] = -0.2, vsini = 1.5km/s, logg* = 4.40cgs. WASP-2 analysis involves SOPHIE RV data with spectroscopic parameters $T_{eff}$ = 5200.0K, [Fe/H] = 0.1 ± 0.2, vsini = km/s, logg* = 4.3cgs \cite{cameron2007wasp}. For HAT-P-30 analysis, we incorporated HIRES and High-Dispersion Spectrograph (HDS; \cite{noguchi2002high}) on the Subaru telescope radial velocity data with spectroscopic parameters $T_{eff}$ = 6304.0 ± 88K, [Fe/H] = -0.13 ± 0.08, vsini = 2.2 ± 0.5km/s, logg* = 4.36 ± 0.03cgs from \cite{johnson2011hat}.

\paragraph{2. Observational data} For this study, we also obtained BVRI photometry data using the 0.8 m remote controlled Ritchey-Chretien telescope equipped with Fingerlake CCD camera. The calibration and target FITS images were obtained at 1x1 binnning with the CCD cooled at $-40^oC$. The observatory is located in Stephenville, Texas, USA at longitude N $32^o$ 12' 58.37" and latitude W $98^o$ 5' 51.46". Further details on the specs of the telescope can be obtained from their website (www.tarleton.edu/observaotry). While the observations were being conducted, the telescope was auto-guided and in focus to gather as much light from the target sources as possible. All data were reduced using standard data reduction routines for bias correction, dark correction subtraction and flat-field correction. Table 03 shows the details of the observations on the three systems.

\begin{table}
\caption{Log of transit observations for three TEP systems presented in this paper where $T_{exp}$ represents observation exposure time for each filter.}\label{tbl1}
    \centering
    \begin{tabular}{ p{3cm} p{3cm} p{3cm} p{3cm} p{2cm} p{1cm} }
\hline
UT Date&     Start Time (UT)& End Time (UT)& No. of Images& $T_{exp}$ (s)& Filter\\
\hline

    &                           &               &
    TrES-3b&          &       \\
    
Aug 16, 2014&           05:07:06&       08:55:27& 
    30&             8&      V\\
   &                           &               &
    30&             5&       R\\
\hline

    &                           &               &
    WASP-2b&          &       \\
    
July 26, 2014&          03:28:05&       07:59:32&
    105&            150&    B\\
   &                           &               &
    105&             15&       V\\
   &                           &               &
    105&             3&       R\\
\hline
    
    &                           &               &
    HAT-P-30b&          &       \\
Feb 08, 2016&           02:04:59&       06:37:30&
    121&            100&    B\\
   &                           &               &
    121&             12&       V\\
   &                           &               &
    121&             0.6&       I\\

\hline
\end{tabular}
\end{table}

\paragraph{3. Data reduction} Differential photometry was carried out using GCX astronomical photometry package on the Linux Mint operating system. Python scripts were also used to implement batch processing procedure to calibrate the raw images and extract Heliocentric Julian Date (HJD) and differential magnitudes to derive the light curves of the three systems. Comparison stars were selected using SIMBAD finder chart and the Tycho catalog. The criteria set for the selection of comparison stars involve: 1)stars must not be variable in brightness, 2) stars should be in vicinity of target star i.e. should not be near the edge of the CCD image, 3) stars must have the same magnitude as of the target star, avoiding their observed stellar flux from being saturated. The apertures were manually assigned on a hit-and-trial basis and we tried a wide range of aperture sizes and retained those which gave photometry with the lowest scatter compared to a fitted model. We extracted light curves using differential photometry by dividing the flux of each comparison star separately with the target star and by taking the average of all the stars, to obtain multiple light curves. The one with the smallest out-of-transit (oot) rms was used as the target’s final light curve for further analysis purposes.

\section{Light Curve Analysis}
\paragraph{1. Light curve modeling} We modeled the light curves extracted from our observational data using publicly available software package: Transit Analysis Package (TAP; \cite{mandel2002analytic}\cite{gazak2012transit}, \cite{carter2009parameter}. TAP utilizes Markov Chain Monte Carlo (MCMC), with metropolis-hasting algorithm and Gibbs sampler and \cite{mandel2002analytic} standard model, derived from a 2-body star-planet system. TAP can determine both temporally correlated (red noise) and uncorrelated noise (white noise). To account for possible temporally correlated noise, TAP uses wavelet likelihood approach of \cite{carter2009parameter} which are more reliable than Chi-squared likelihood techniques (CW09, \cite{johnson2012characterizing}.

\paragraph{} By using TAP, following parameters can be fitted simultaneously: orbital period ($P_b$), mid transit time ($T_m$), orbital inclination (i), scaled semi-major axis (a/R*), planet-to-star radii ratio ($R_p$/R*), linear and quadratic limb darkening coefficients ($u_1$ and $u_2$), orbital eccentricity (e) and longitude of periastron (w). Light curve data uploaded to the TAP once the light curve flux is normalized in order to make out-of-transit values (Oot) = 1. We modelled transit light curves with MCMC using 10 chains with the length of $10^4$ links each. Before fitting the extracted light curves, initial values are assigned to the parameters followed by decision of which parameters needed to be free and fixed. During analysis, the inclination (i), time of mid transit ($T_c$), ratio of planet-to-star radius ($R_p$/R*) and semi-major axis to star radius ratio (a/R*) were set as free parameters. Eccentricity (e), argument of periastron (w), linear and quadratic limb darkening coefficients ($u_1$ and $u_2$) and the planet’s orbital period (P) were fixed. Value of eccentricity (e) and argument of periastron (w) can be determined by combining the data acquired from the planet's primary eclipse and radial velocity curve or combined secondary eclipse calculation with the primary eclipse. As planets are hot Jupiters which have very short orbital periods and also exhibit synchronous rotations with respect to their host stars giving rise to the tidal effects. We assumed TrES-3 system planet’s orbit is circular as suggested by \cite{o2007tres}, \cite{fressin2010broadband}(p. 3) and the period was set to P = 1.306186days \cite{christiansen2010system}. Similarly, the HAT-P-3b system, \cite{enoch2011wasp} found its eccentricity = 0. A complete description of parameters fitted in TAP is given in Table 4.

\paragraph{} While fitting the transit light curves, limb darkening coefficients play an important role. The values of limb darkening coefficients vary according to the filters being used during observations. In this perspective, linear ($u_1$) and quadratic ($u_2$) limb darkening coefficients in each respective band were acquired from \cite{claret2004irradiated}\cite{claret2012new} and interpolated to the effective temperature ($T_{eff}$), metallicity ([Fe/H]) and surface gravity (logg*) of candidate stars as given in table 2. Values of limb darkening coefficients obtained for each band are listed in Table 6. 

\paragraph{} We also accounted for the temporally uncorrelated Gaussian (white noise) and temporally correlated Gaussian (red noise) and were set as free parameters in TAP. An important fact to consider when fitting light curves is red noise; as neglecting this factor can undervalue the errors and uncertainties and hence can result in incorrect best-fitting parameter values \cite{pont2006effect} (CW09; \cite{gazak2012transit}).

\begin{table}
\caption{Parameter setting for light curve fitting in TAP software. Initial values of $P_b$, i, a/R*, $R_p$/R* for TrES-3b, WASP-2b and HAT-P-30b are set as the values in \citet{sozzetti2009new}, \citet{cameron2007wasp} and \citet{johnson2011hat}}\label{tbl1}
    \centering
    \begin{tabular}{p{2cm} | p{2cm} p{2cm} | p{2cm} p{2cm} | p{2cm} p{2cm}} 
\hline
 &   \multicolumn{2}{|c|}{TrES-3 System}&    \multicolumn{2}{c|}{WASP-2 System}&      \multicolumn{2}{c}{HAT-P-30 System}\\
\hline
Parameters& Initial Values& During MCMC Chains& Initial Values& During MCMC Chains&  Initial Values& During MCMC Chains\\
\hline

$P_b$ (days)&           1.30618581&             Fixed&
                    2.152226&             Fixed&                  2.810595&               Fixed\\

i (deg)&           81.85&             Free&
                    84.74&             Free&             
                    83.6&             Free\\

a/R*&           5.926&             Free&
                8.8495&             Free&                  
                7.42&              Free\\

$R_p$/R*&       0.1655&             Free&
                0.13&             Free&                  
                0.1134&             Free\\

$T_m$&           &             Free&
              &             Free&                  &
Free\\

$u_1$&           &             Fixed&
              &             Fixed&                  &
Fixed\\

$u_2$&           &             Fixed&
              &             Fixed&                  &
Fixed\\

$e$&           0&             Fixed&
              0&             Fixed&                  0&
Fixed\\

$w$&           0&             Fixed&
              0&             Fixed&                  0&
Fixed\\

Source papers&  \cite{sozzetti2009new}&             &
             \cite{cameron2007wasp}&             & \cite{johnson2011hat}&              \\

\hline
\end{tabular}
\end{table}

\paragraph{} Moreover, we found that when the deviations of transit-light-curve data are smaller, the resulting error bars become smaller. Thus, the error bar does reflect the quality of data. The MCMC procedure in TAP gives a reasonable estimation on the error related to the data itself. Therefore, the error bars we obtained here should have been consistent with the scattering and quality of the light curves, and provide reliable error estimates. For TrES-3b, WASP-2b and HAT-P-30b, the best-fitted parameters values of $T_m$, i, $R_p$/R*, a/R*, limb darkening coefficients ($u_1$, $u_2$) obtained from TAP, along with previously calculated values and the derived transit durations are shown in table 7, 8, 9 and 5 respectively. Moreover, the observed light curves and best fitting models of our own data are presented in Figure 1 through 8, where the points are observational data and solid curves are the best fitting models. The transit duration ($t_au$) of each of our transit model fits is calculated using following equation from \cite{carter2008analytic}:\\
($t_{au}$) = $t_{egress}$ – $t_{ingress}$

\paragraph{2. Period determination} We can improve the ephemeris for TrES-3b, WASP-2b and HAT-P-30b by using our TAP calculated mid-transit times and comparing them with previously published mid-transit times, to refine the orbital period of the planet and looking for any transit timing variations. Mid-transit times calculated in this study are summarized in table 7. 

\paragraph{} For this, the Julian Date mid-transit times were transformed into BJD which is based on TDB using the online converter by \cite{eastman2010achieving}. We calculated the new refined ephemeris by using the following equation:\\
$T_mc$ = $T_0$ + ($P_b$)(E) \\
where $P_b$ is the orbital period of the planet, E is the integer number of cycles (primary eclipse epoch) after the discovery paper and $T_0$ is the reference time when the planet was discovered for which generally epoch E = 0; $T_mc$ is mid-transit time calculated under E = 0. All the derived values are written in table 07, table 08 and table 09 for TrES-3b, WASP-2b and HAT-P-30b respectively.\\

\section{Physical properties of the systems}

\paragraph{} We redetermined the physical properties of TrES-3, WASP-2 and HAT-P-30 planetary systems by combining together the measured parameters from our light curve modeling (a/R*, $R_p$/R*, $T_m$, i, $P_b$) and spectroscopic data from referenced sources as given in table 02. We used the results obtained from light curve modelling with TAP along with calculations from previous literature to estimate the planetary and geometrical properties (i.e. mass, radius, density, surface gravity, equilibrium temperature, Safronov number, atmospheric scale height). The physical parameters of all the three targets can be found in table 10. We adopted the equation by \cite{winn2010hot} \cite{seager2011characterizing} to find the planetary mass,

 \begin{ceqn}
\begin{align}
 M_{b}=\left(\frac{\sqrt{1-e^{2}}}{28.4329}\right)\left(\frac{\gamma}{\sin i}\right)\left(\frac{P_{b}}{1 y r}\right)^{1 / 3}\left(\frac{M_{*}}{M_{\odot}}\right)^{2 / 3} M_{\text {jup }}
 \end{align}
\end{ceqn}

where $K_*$ is the radial velocity semi-amplitude, $P_b$ is the orbital period of the planet.

\paragraph{} The surface gravitational acceleration, $g_{b}$, can be calculated using the following formula by \cite{southworth2007method}:

\begin{ceqn}
\begin{align}
g_{b}=\frac{2 \pi}{P_{b}}\left(\frac{a}{R_{b}}\right)^{2} \frac{\sqrt{1-e^{2}}}{\sin i} K
 \end{align}
\end{ceqn}

Where $K_{∗}$ is the stellar velocity amplitude listed in Table 9.
The equilibrium temperature, $T_eq$, was derived using the relation \cite{southworth2010homogeneous}

\begin{ceqn}
\begin{align}
T_{eq}=T_{eff} \sqrt{\left(\frac{R *}{2 a}\right)}
\end{align}
\end{ceqn}

where $T_{eff}$ is the effective temperature of the host star.
We calculated the Safronov number, $\Theta$, using the equation from \cite{southworth2010homogeneous}:

\begin{ceqn}
\begin{align}
\Theta=\frac{M_{b} a}{M_{s} R_{b}}
\end{align}
\end{ceqn}

\begin{table}
\caption{Photometric light curve data for this work.}\label{tbl1}
    \centering
    \begin{tabular}{p{2cm} p{2cm} p{2cm} p{4cm} p{4cm}} 
\hline
 Name& UT Date& Filter&  TDB-based BJD&  Relative Flux\\
\hline
        &           &   &   2456885.75031372&   1.00285928519769\\
        &           &  V&   2456885.79510193&   0.98300000000000\\
TrES-3b& 2014 Aug 16&   &   2456885.82162087&   0.99641441058989\\
        &           &\\\hline   &              &                   \\
        &           &   &   2456885. 7132652&   0.999079389837209\\
        &           &  R&   2456885.79561191&   0.984917830835652\\
        &           &   &   2456885.82213085&   0.998159627344608\\
\hline
        &           &   &   2456864.64450919&   1.005314943817520\\
        &           &  B&   2456864.73294038&   0.976637359760591\\
        &           &   &   2456864.83205170&   0.999079390445019\\
        &           &\\\hline   &              &                   \\
        &           &   &   2456864.65227930&   1.002143699400000\\
WASP-2b& 2014 Jul 26&  V&   2456864.73530041&   0.981182966900000\\
        &           &   &   2456864.83262171&   0.996207943500000\\
        &           &\\\hline   &              &                   \\
        &           &   &   2456864.64546921&   1.00018422438524\\
        &           &  R&   2456864.73930047&   0.977867472820088\\
        &           &   &   2456864.83301171&   1.000092108256000\\
\hline
        &           &   &   2457426.59731026&   1.001536235548020\\
        &           &  B&   2457426.68499830&   0.978137704339269\\
        &           &   &   2457426.78338610&   0.999079390001093\\
        &           &\\\hline   &              &                   \\ &           &   &   2457426.62725959&   1.001382825000000\\
HAT-P-30b&2016 Feb 08&  V&  2457426.69463809&   0.986279485474359\\
        &           &   &   2457426.76619648&   0.999999999840966\\
        &           &\\\hline   &              &                   \\
        &           &   &   2457426.60117017&   1.000000000046010\\
        &           &  I&   2457426.68436832&   0.983558056272207\\
        &           &   &   2457426.77604626&   0.998619402892466\\

\hline
\end{tabular}
\end{table}

\begin{table}
\caption{Theoretical limb darkening coefficients ($u_1$, $u_2$) for the TrES-3, WASP-2 and HAT-P-30 stars.}\label{tbl1}
    \centering
    \begin{tabular}{p{2cm} | p{2cm} p{2cm} | p{2cm} p{2cm} | p{2cm} p{2cm}} 
\hline
 Filters&   \multicolumn{2}{|c|}{TrES-3b}& \multicolumn{2}{c|}{WASP-2b}&      \multicolumn{2}{c}{HAT-P-30b}\\
\hline
 &          u1&             u2&             u1&             u2&                u1&             u2\\
\hline
B&           -&              -&          0.6758&        0.1143&            0.7932&          0.1131\\

V&      0.4378&         0.2933&          0.5405&        0.0350&             0.4916&          0.2409\\

R&      0.567&         0.3178&          0.6674&        0.2091&                  -&              -\\

I&          -&              -&               -&             -&              0.8861&         0.0675\\

\hline
\end{tabular}
\end{table}

\paragraph{} The Safronov number is a measure of the ability of a planet to gravitationally scatter or capture other bodies \cite{safronov1972ejection}. Difference values of the Safronov number indicate different probabilities in migration or stopping mechanism of planetary bodies. According to which, planets with Safronov number, $\Theta$ = 0.07±0.01 classified as class I and class II planets have Safronov number, $\Theta$ = 0.04±0.01 as defined by \citet{hansen2007two}. 

Atmospheric scale height, H, is calculated using \cite{de2013constraining}:

\begin{ceqn}
\begin{align}
H=\frac{k_{\mathrm{B}} T_{\mathrm{eq}}}{\mu_{\mathrm{m}} g_{\mathrm{p}}}
\end{align}
\end{ceqn}

$k_B$ is the Boltzmann constant and $\mu$ is the mean molecular weight in the planet's atmosphere. Resulting estimates of physical parameters of each planetary system are listed in the table 10.

\begin{table}
\caption{Parameter values for TrES-3b derived from light curve fitting for $T_m$, i,  a/R*, $R*_p$/R*, $u_1$ and $u_2$. Data sources (a):\citet{sozzetti2009new}, (b):\citet{gibson2009transit}, (c):\citet{colon2010characterizing}, (d):\citet{jiang2013possible}, (e):\citet{christiansen2010system} and (f):\citet{turner2013near}}\label{tbl1}
    \centering
    \begin{tabular}{ c | c | c | c | c | c | c | c } 
\hline
 Epoch& $T_m$& i&  a/R*&  $R*_p$/R*& $u_1$& $u_2$& Data Source\\
\hline

    0&  4185.911±0.00021&     81.90±0.11&     5.906±0.044&    
    0.1656±0.0024&    0.219±0.049&    0.317±0.050&    (a)\\

    10& 4198.97359±0.0006&    81.79±0.14&    5.944±0.052&
    0.1734±0.0072&    0.642±0.049&      0.183±0.049&  (a)\\
    
    22& 4214.64695±0.00034&   81.81±0.12&    5.915±0.046&
    0.1708±0.0045&     0.440±0.049&      0.292±0.049& (a)\\
    
    23& 4215.95288±0.00032&    81.77±0.13&   5.937±0.047&
    0.1716±0.0054&    0.563±0.049&      0.242±0.049&  (a)\\
    
    267& 4534.66317±0.00019&  81.79±0.12&   5.952±0.044&
    0.1691±0.0038&    0.393±0.05&       0.312±0.05&   (b)\\
    
    268& 4535.96903±0.00038&  81.83±0.12&   5.920±0.048&
    0.1639±0.0043&    0.283±0.05&       0.322±0.049&  (a)\\
    
    270& 4538.58069±0.00020&  81.99±0.30&   6.01±0.0204&
    0.1661±0.03&      0.5169&             −0.6008& (e)\\
    
    281& 4552.94962±0.00020&  81.86±0.11&   5.939±0.043&
    0.1634±0.0033&    0.350±0.049&      0.308±0.050&  (a)\\
    
    294& 4569.92982±0.0004&   81.82±0.12&    5.926±0.047&
    0.1648±0.0059&     0.275±0.049&      0.317±0.050&  (a)\\
    
    313& 4594.74682±0.00037&   81.85±0.12&    5.943±0.047&
    0.1646±0.005&     0.273±0.05&       0.317±0.05&   (a)\\
    
    329& 4615.64621±0.0002&   81.81±0.11&   5.915±0.044&
    0.1674±0.003&     0.4±0.05&         0.313±0.049&  (b)\\

    342& 4632.6269±0.0002&    81.81±0.11&   5.937±0.042&
    0.1685±0.0036&    0.39±0.048&       0.307±0.049&  (b)\\
    
    355& 4649.60712±0.0002&   81.81±0.11&   5.931±0.042&
    0.1649±0.0030&    0.381±0.048&      0.297±0.049&  (b)\\
    
    358& 4653.52661±0.00091&  81.89±0.14&   5.913±0.052&
    0.1676±0.008&     0.385±0.049&      0.302±0.05&   (b)\\
    
    365& 4662.66984±0.0006&   81.90±0.13&   5.915±0.051&
    0.1664±0.007&     0.386±0.05&       0.303±0.05&   (b)\\
    
    371& 4670.50709±0.00034&  81.86±0.11&   5.885±0.046&
    0.1619±0.003&      0.393±0.048&      0.305±0.049&  (b)\\
    
    374& 4674.42521±0.00028&  81.74±0.11&    5.965±0.047&
    0.1616±0.004&     0.390±0.049&      0.308±0.049&  (b)\\
    
    381& 4683.56812±0.00042&  81.85±0.12&   5.927±0.048&
    0.1644±0.005&     0.387±0.05&       0.305±0.05&   (b)\\
    
    620& 4995.75025±0.0005&    79.9±1.1&     5.36±0.25&
    0.214±0.05&                   -&                   -&   (f)\\
    
    627& 5004.89249±0.0002&    80.9±0.9&     5.68±0.29&
    0.188±0.03&                   -&                   -&   (f)\\
    
    637& 5017.95452±0.0004&    82.0±0.6&      6.02±0.32&
    0.166±0.02&                   -&                   -&   (f)\\
    
    665& 5054.52523±0.00018&   81.83±0.1&     5.932±0.043&
    0.1655±0.0026&     0.260±0.049&      0.320±0.049&   (c)\\
    
    885& 5341.8838±0.001&      81.81±0.15&   5.935±0.053&
    0.1683±0.009&      0.345±0.05&        0.323±0.05&   (d)\\
    
    898& 5358.86606±0.00076&   81.75±0.14&   5.957±0.053&
    0.1627±0.009&     0.345±0.05&       0.324±0.05&   (d)\\
    
    901& 5362.7847±0.0011&    81.83±0.15&   5.937±0.053&
    0.160±0.011&      0.341±0.05&       0.320±0.05&   (d)\\
    
    904& 5366.70215±0.0008&   81.95±0.14&   5.987±0.052&
    0.1646±0.007&      0.334±0.05&       0.296±0.05&   (d)\\
    
    911& 5375.84617±0.0009&   81.89±0.14&   5.912±0.052&
    0.1543±0.008&     0.338±0.05&       0.316±0.05&   (d)\\
    
    1273& 5848.6873±0.0011&   84.1±2.1&      7.10±1.30&
    0.146±0.030&                   -&                   -&   (f)\\
    
    1289& 5869.5836±0.0017&   79.4±2.0&     5.18±1.06&
    0.227±0.05&                   -&                   -&   (f)\\
    
    1398& 6011.9597±0.00081&  80.9±1.5&     5.85±0.4&
    0.208±0.06&                   -&                   -&   (f)\\
    
    1400& 6014.57295±0.00073& 79.7±1.2&     5.36±0.3&
    0.227±0.05&                    -&                   -&   (f)\\
    
    1411& 6028.94120±0.0005&   82.4±0.8&     6.14±0.41&
    0.153±0.01&                   -&                   -&   (f)\\
    
    3067& 6885.79665±0.00008&  81.85&           6.731±0.051&
    0.1582±0.005&              0.369&          0.348& This work\\

\hline
\end{tabular}
\end{table}

\begin{table}
\caption{Parameter values for WASP-2b derived from light curve fitting for $T_m$, i,  a/R*, $R*_p$/R*, $u_1$ and $u_2$. Data sources (a): \citet{cameron2007wasp}, (b): \citet{sada2012extrasolar}, (c): \citet{turner2017investigating}}\label{tbl1}
    \centering
    \begin{tabular}{ c | c | c | c | c | c | c | c } 
\hline
 Epoch& $T_m$& i&  $a/R*$&  $R*_p$& $u_1$& $u_2$& Data Source\\
\hline

    0&  3991.5146±0.0044&              &     11.6279-7.5757&    0.119-0.140&            &                   &       (a)\\

    356& 4757.70492±0.00032&    85.18±1.26&      8.22±1.08&
    0.1135±0.0053&     0.170&              0.341&          (b)\\
    
    1316& 6823.83839±0.00055&  84.86±1.61&    8.05±1.21&
    0.1383±0.0049&      0.8227&         0.0186&    (c)\\
    
    1335& 6864.73015±0.00016&   85.07&          8.10±0.25&
    0.1324&             0.6279&             0.1194&     This work\\

\hline
\end{tabular}
\end{table}

\section{Discussion}

\subsection{Individual Systems}

\paragraph{}\emph{1. TrES-3b}
\paragraph{} To derive the physical properties of the TrES-3 system as shown in the table, values of physical parameters are obtained from \cite{sozzetti2009new}. Our derived results are consistent with the previous studies published by \citet{o2007tres}, \citet{southworth2010homogeneous}, \citet{lee2011physical} and \citet{turner2013near} with a small difference in planetary mass
values. From our analysis, we deduced that the planet has a mass ($M_p$) of 1.773$M_{jup}$ that is smaller in comparison to earlier findings. Likewise, planetary radius is found to be $R_p$ = 1.305$R_{jup}$ resulting in density, $\rho_{p}$ = 0.7982$\rho_{jup}$. TrES-3b appears to have a constant planetary radius in optical (V and R) wavelength filters used for the observations. TrES-3b is almost grazing its host star by looking the shape of its light curves obtained. There must be a minimum angle of inclination ($i_{gr}$), below which planet would likely be grazing it's host star, i.e. part of planet's disk will not be inside stellar disk even when planet is fully in front of it's host star. Following equation provides the theoretical limit also discussed by \citet{hartman2019hats}:

\begin{ceqn}
\begin{align}
i_{\mathrm{gr}}=\frac{R_{\mathrm{*}}-R_{\mathrm{p}} }{a}
\end{align}
\end{ceqn}

For TrES-3b, this inclination limit is $i_{gr}$ = 82.80. This is obtained by using the planetary and stellar radii and semi-major axis values from table 01 and table 10. The orbital inclination of TreS-3b determined by fitting its light curves using TAP modeling package is 81.85. This shows that transit is almost grazing, (also confirming the planetary radius values obtained from modeling the transit light curves of TrES-3b). Therefore, we only relied on theoretical limb darkening coefficients and did not allow them to fit them in our light curve modeling with TAP. Equilibrium temperature ($T_{eq}$) determined in our calculation is also lower than preceding values with $T_{eq}$ = 1554.5623K. It is also observed that the radius of the planet is almost the same in different filters from visible to I bandpass wavelengths as determined by \citet{turner2013near}.

\begin{table}
\caption{Parameter values for HAT-P-30b derived from light curve fitting for $T_m$, i,  $a/R*$, $R*_p/R*$, $u_1$ and $u_2$. Data sources (a): \citet{johnson2011hat}, (b): \citet{enoch2011wasp} and (c): \citet{maciejewski2016new}}\label{tbl1}
    \centering
    \begin{tabular}{ c | c | c | c | c | c | c | c } 
\hline
 Epoch& $T_m$& i&  a/R*&  $R*_p$& $u_1$& $u_2$& Data Source\\
\hline

    0& 5456.46561±0.00037&      83.6±0.4&       7.42±0.26&
    0.1134±0.0020&      0.1975&             0.3689&     (a)\\
    
    42& 5574.51188±0.00050&    82.48±0.15&    6.67±0.17&
    0.1095&                 &                   &       (b)\\
    
    174& 5945.51205±0.00052&   82.70±0.19&     6.771±0.013&
    0.1109±0.0015&         &                   &   (c)\\
    
    701& 7426.69346±0.0003&    83.79&          6.916&
    0.125&              0.6971&         0.1354&         This work\\

\hline
\end{tabular}
\end{table}

\begin{table}
\caption{Estimated weighted-mean parameter values for TrES-3, WASP-2 and HAT-P-30 systems.}
    \centering
    \begin{tabular}{ p{3cm} | p{3cm} | p{3cm} | p{3cm} } 
\hline
    Planet&         TrES-3b&    WASP-2b&    HAT-P-30b\\
\hline

    $M_b$ ($M_{jup}$)&      1.7730&     0.8374&     0.7006\\
    $R_b$ ($R_{jup}$)&      1.3047&     1.0274&     1.5109\\
    $\rho_b$($\rho_{jup}$)&      0.7983&     0.7721&     0.2031\\
    log$g_b$ (cgs)&   3.55&       3.3468&     2.8463\\
    $T_{eq}$(K)&       1554.5623&  1253.1871&  1695.0156\\
    H(km)&         248&           382&           1450\\
    $\Theta$&              0.076&      0.0657&     0.029\\
    i&              81.85&      85.07&      84.79\\
    a(AU)&         0.02594&    0.02937&    0.03906\\
    $T_{m}$& (BJD) 6885.7956&  6864.7353& 7426.6946\\
\hline
\end{tabular}
\end{table}

\paragraph{}\emph{2. WASP-2b}
\paragraph{} For the WASP-2b system, spectroscopic data was obtained from \cite{cameron2007wasp}. Our derived physical parameters agree with the previous literature, with planetary radius ($R_p$) = 1.027 $R_{jup}$ as determined by \citet{charbonneau2007precise}, \citet{southworth2010homogeneous}, \citet{turner2016vizier} and $M_p$ = 0.8375 $M_{jup}$ in consistent with values calculated by \citet{cameron2007wasp}, \citet{southworth2010homogeneous} and \citet{triaud2010spin}. WASP-2b has a density, $\rho_p$ = 0.7721$\rho_{jup}$ as calculated by \citet{southworth2009homogeneous}, \citet{southworth2010homogeneous} and \citet{turner2017investigating}. WASP-2b is orbiting around its host star at a distance of 0.02937 AU in a circular orbit, with an equilibrium temperature, $T_{eq}$ = 1253.1871K. All the estimated physical parameters are listed in the table.

\paragraph{}\emph{5.3 HAT-P-30b/WASP-51b}
\paragraph{} With our new transit light curves, and after fitting them with appropriate models in TAP, we were able to extract the physical properties of the HAT-P-30 system. Our derived parameters estimates planetary mass ($M_p$) of 0.7006 Mjup and a bloated planetary radius ($R_p$) of 1.5109 Rjup, diverging from previous studies by \citet{johnson2011hat}, \citet{enoch2011wasp} and \citet{maciejewski2016new}, put HAT-P-30b into an inflated short-period hot Jupiter, with the estimated equilibrium temperature, $T_{eq}$ = 1695.0156K and a comparatively low density, $\rho_{p}$ = 0.2031$\rho_{jup}$, determined from the values of our derived planetary mass and radius parameters. Hence, there is a low surface gravitational acceleration $g_p$ = 7.0195 $ms^{-2}$ as compared to previous studies, is calculated from our derived values.

\subsection{Variations in the planetary radii w.r.t wavelength}
\paragraph{} Obtaining photometric light curves in different filters can be used to detect any changes in the radius of the planet in each bandpass, hence approximating the chemical composition in their atmospheres. For this purpose, we used our multi-color photometric observations to probe the possible variations in the radii of TrES-3b, WASP-2b and HAT-P-30b in different optical filters i.e B,V,R and I passbands \cite{ciceri2015physical}.

\paragraph{} We observed a constant transit depths across optical wavelengths for the TEPs TrES-3b and WASP-2b (except of one value of WASP-2b) (see from figure 1 through 5). Our $R_p$/$R^*$ for R-band observed for WASP-2b is found to be different from previously determined $R_p$/$R^*$ value by 0.14$\sigma$. The reason is unclear and future precise observations are required to investigate the possible reasons. The relatively flat spectrum in the radius of these two TEPs suggest that these planets might have clouds or haze in their upper atmospheres \cite{seager2000theoretical}, \cite{brown2001transmission}, \cite{gibson2013optical}, \cite{marley2013clouds} and \cite{kreidberg2014clouds} or there is also a possibility of presence of isothermal pressure-temperature profiles \cite{fortney2006influence}. However, we find variations in the transit depths with wavelength for HAT-P-30b (see figure 06 to 08). Radius variations in HAT-P-30b show significantly large transit depth in blue filter than rest of the optical band-pass. \citet{evans2016detection} maintains that possible reason for the variation can be the TiO/VO absorptions in the atmosphere. This may also indicate the variation in the particle size in the high altitudes of its atmosphere. Nevertheless, this increase in the planetary size put it into the category of bloated/inflated TEPs. The large value of scale height (H) determined for HAT-P30b also confirms our large planetary radius (see table 10). 

\section{Conclusion}
\paragraph{} Using Tarleton State University Observatory telescope, we analyzed new multi-color photometric light curves of transit events in TrES-3, WASP-2 and HAT-P-30, and reported their refined parameters. We modeled the light curves using the TAP package and the main physical parameters of the planetary systems. Our results are generally in agreement with previously published data with the exception of a few parameters that lean towards small and less massive planets. i.e TrES 3b appears to be a smaller and less-massive planet with the following  parameters of radius, $R_b$ = 1.305$R_{jup}$ and mass $M_{b}$ = 1.773$M_{jup}$. HAT-P-30 appears to be a inflated planet with the following parameters, $R_{b}$ = 1.5109$R_{jup}$ and $M_{b}$ = 0.7006$M_{jup}$ (see table 10). We used these new photometric multi-band observations to obtain wavelength-dependent measurement of ratio of planet to star radius in optical windows and some near-IR windows. We measured R-band radius ratio higher than the other bands, but with a non-significant level of confidence.\\
1 meter class telescopes can obtain high precision multi-color transit photometry data that enable accurate astrophysical modeling and analysis. Multi-color photometry light curves allow us to derive the variation in radius of the planet as a function of wavelength and compare the results with existing published data. 
Analysis show that the radii are constant in the visible to near-IR region. We believe that the small variations in parameters derived from TAP is due to either the stratospheric composition of transiting planet or due to systematic  differences in limb darkening coefficient values that were adopted from \citet{claret2013new} and \citet{mackebrandt2017transmission}. Interestingly \citet{christiansen2010system} also propose long term stellar variability in TrES-3b as also as one of the possibilities of this variation.

Combining our data with previous studies, systems appear to have constant radius through visible to near-IR. We observed small variations in parameters derived from TAP. This can occur due to two possibilities, either due to atmospheric composition of transiting planets or offset might be caused by the systematic differences in limb darkening coefficient values taken from  \citet{claret2013new} as also mentioned in \citet{mackebrandt2017transmission}. It can also be due to long-term stellar variability in case of TrES-3 as discussed by \citet{christiansen2010system}.

\section{Acknowledgements}
\paragraph{} We are extremely thankful to Tarleton State University (TSU) for providing us with the remote access to their observatory to collect data from their telescope, allowing us to perform data reduction and analysis processes and also providing us access to their other computational facilities.

\clearpage

\paragraph{} 

\begin{figure}
	\centering
		\includegraphics[scale=.5]{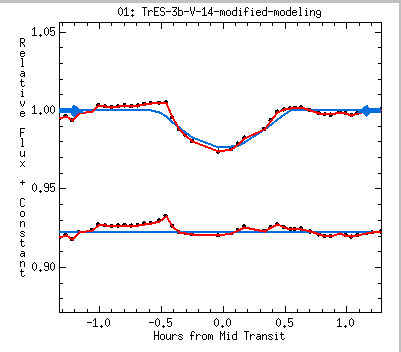}
	\caption{Modelled individual transit curves of TrES-3b in V- filter with photometric error bars and residuals from fits. The normalized relative flux as a function of the time. Points are the data and red-line curves are the best-fitting models obtained from TAP.}
	\label{FIG:1}
\end{figure}

\begin{figure}
	\centering
		\includegraphics[scale=.5]{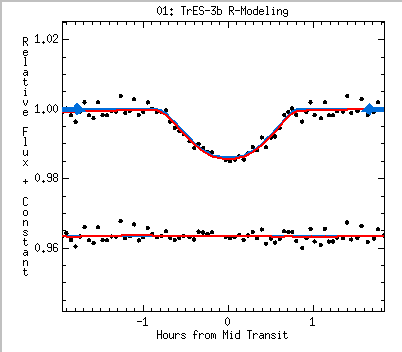}
	\caption{Modelled individual transit curves of TrES-3b in R- filter with photometric error bars and residuals from fits. The normalized relative flux as a function of the time. Points are the data and red-line curves are the best-fitting models obtained from TAP.}
	\label{FIG:1}
\end{figure}

\begin{figure}
	\centering
		\includegraphics[scale=.5]{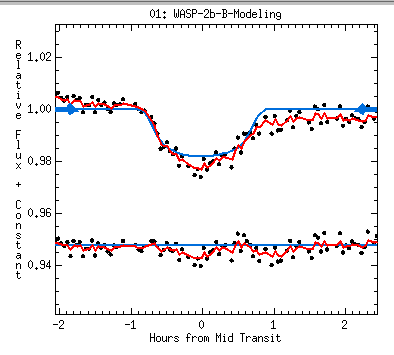}
	\caption{Modelled individual transit curves of WASP-2b in B- filter with photometric error bars and residuals from fits. The normalized relative flux as a function of the time. Points are the data and red-line curves are the best-fitting models obtained from TAP.}
	\label{FIG:1}
\end{figure}

\begin{figure}
	\centering
		\includegraphics[scale=.5]{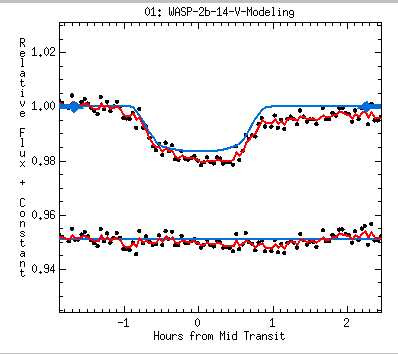}
	\caption{Modelled individual transit curves of WASP-2b in V- filter with photometric error bars and residuals from fits. The normalized relative flux as a function of the time. Points are the data and red-line curves are the best-fitting models obtained from TAP.}
	\label{FIG:1}
\end{figure}

\begin{figure}
	\centering
		\includegraphics[scale=.5]{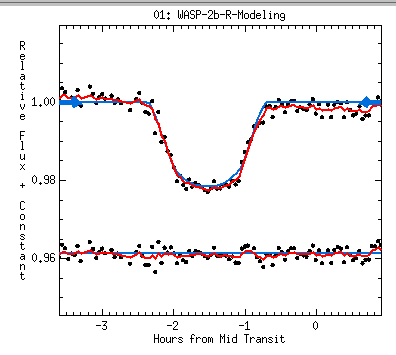}
	\caption{Modelled individual transit curves of WASP-2b in R- filter with photometric error bars and residuals from fits. The normalized relative flux as a function of the time. Points are the data and red-line curves are the best-fitting models obtained from TAP.}
	\label{FIG:1}
\end{figure}

\begin{figure}
	\centering
		\includegraphics[scale=.5]{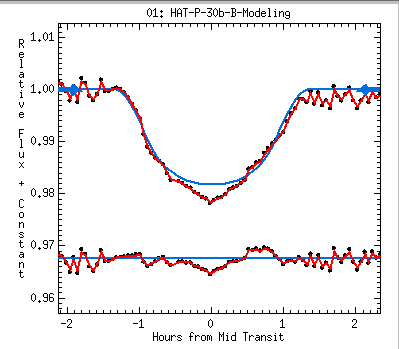}
	\caption{Modelled individual transit curves of HAT-P-30b in B- filter with photometric error bars and residuals from fits. The normalized relative flux as a function of the time. Points are the data and red-line curves are the best-fitting models obtained from TAP.}
	\label{FIG:1}
\end{figure}

\begin{figure}
	\centering
		\includegraphics[scale=.5]{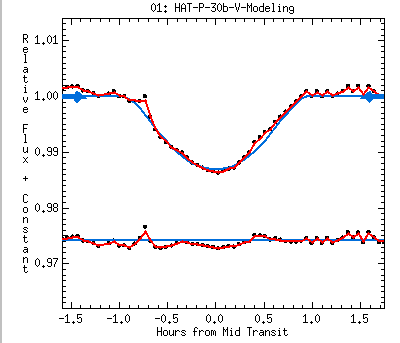}
	\caption{Modelled individual transit curves of HAT-P-30b in V- filter with photometric error bars and residuals from fits. The normalized relative flux as a function of the time. Points are the data and red-line curves are the best-fitting models obtained from TAP.}
	\label{FIG:1}
\end{figure}

\begin{figure}
	\centering
		\includegraphics[scale=.5]{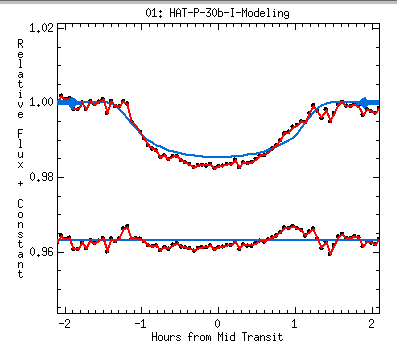}
	\caption{Modelled individual transit curves of HAT-P-30b in I- filter with photometric error bars and residuals from fits. The normalized relative flux as a function of the time. Points are the data and red-line curves are the best-fitting models obtained from TAP.}
	\label{FIG:1}
\end{figure}

\clearpage

\clearpage

\bibliographystyle{cas-model2-names}

\bibliography{ExoBib}


\end{document}